\documentstyle[floats,aps,epsf]{revtex}
\begin{document}
\twocolumn[\hsize\textwidth\columnwidth\hsize\csname@twocolumnfalse%
\endcsname
\draft

\title{Fluid adsorption at a non-planar wall:\\ roughness induced first-order wetting}
\author{A.\ O.\ Parry, P.\ S.\ Swain and J.\ A.\ Fox}
\address{Department of Mathematics, Imperial College \\ 180 Queen's Gate, London SW7 2BZ, United Kingdom.}
\maketitle

\begin{abstract}
We study the problem of fluid adsorption at a non-planar wall with a view to understanding the influence of surface roughness on the wetting transition. Starting from an appropriate Landau-type free energy functional we develop a linear response theory relating the free energy of the non-planar system to the correlation functions in its planar counterpart. Using this approach we are able to generalize the well known graphical construction method used to study the planar surface phase diagram and derive analytical expressions for the shift in the phase boundary for first and second-order wetting transitions. The results of the calculation are compared and contrasted with simple phenomenological and scaling arguments. Of particular interest is the influence of surface roughness on a second order wetting transition which is driven first order, even for small deviations from the plane.
\end{abstract}

\vskip2pc]

While the statistical mechanical theory of fluid adsorption at planar walls and in other idealized geometries (such as capillary-slits and cylindrical-pores) is much studied \cite{RW,henderson} the microscopic theory of fluid adsorption at non-planar (rough) walls is far less developed. Nevertheless this problem is certainly of practical interest as well as posing something of a theoretical challenge due to the loss of translational invariance. For adsorption at a single wall perhaps the most important issue is whether the roughness influences the order and location (phase boundary) of any wetting transition \cite{schick}. Here we address this question using a generalized Landau density-functional model which may be viewed as the simplest available microscopic approach \cite{H}. Recall that in application to theories of the planar liquid-vapour interface as well as wetting transitions and finite-size effects the same approach has played a pivotal role \cite{fisher} leading to more sophisticated methods. It is therefore a natural starting point for the systematic investigation of non-planar fluid interfaces. A major part of our work is to show that within perturbation theory (valid for small deviations from the plane) it is possible to derive an analytical expression for the change in free energy due to roughness that may be studied using a generalization of a graphical construction method familiar from the planar problem \cite{sullivan}. In this way we avoid having to use effective Hamiltonian methods \cite{rejmer}. Our analysis borrows results and methods recently developed for the calculation of pair correlation functions at planar wall-fluid interfaces \cite{parry} and is we believe, of pedagogical interest beyind its application reported here. Using this approach we are able to derive analytical expressions for a shift of the phase boundary at first and second-order wetting transitions in three dimensional systems. While our analysis is mean field-like we are confident that the topology of the surface phase diagram is uneffected by the inclusion of fluctuation effects.
	
To begin we make some preliminary remarks concerning the possibility of a roughness induced wetting temperature shift. It was observed by Wenzel \cite{wenzel} some sixty years ago that the contact angle $\theta_\Xi$ of a drop of planar liquid (phase $\beta$ say) on a rough substrate (wall-$\alpha$ phase interface) appeared to satisfy the empirical relation
\begin{equation}
\cos \theta_\Xi = r \cos \theta_\pi
\end{equation}
where $r \equiv \frac{A_\Xi}{A_\pi}$ is the ratio of non-planar to planar surface area and $\theta_\pi$ is the contact angle for the planar system ($r=1$). Of course this is not an exact result but has nevertheless been proved (rigorously) valid for an Ising model with a non-planar boundary at sufficiently low temperatures \cite{borgs}. Unfortunately the low temperature restriction precludes a study of the influence of roughness on any wetting transition. However it is clear that if Wenzel's law was valid $\forall \theta_\Xi \ge 0$ then the wetting temperature is necessarily reduced. This is conveniently expressed as follows : the non-planar wall-$\alpha$ interface is completely wet (i.\ e.\ $\theta_\Xi=0$) by a fluid $\beta$ phase for roughness parameter $r \ge r_W$ where
\begin{equation}
r_W = \sec \theta_\pi \label{wenzel1}
\end{equation}
and recall $\theta_\pi$ is the contact angle ($>0$) of the $\beta$ droplet at the planar wall-$\alpha$ interface (at two phase bulk coexistence). If we assume that the wall has a corrugated shape described by the graph $z_W({\bf r}_\parallel) = \sqrt{2} a \sin q x$ where ${\bf r}_\parallel=(x,y)$ is the displacement vector parallel to the $z=0$ plane (corresponding to the mean position of the wall) then we can rewrite the equation for the phase boundary as
\begin{equation}
\theta_\pi = q a \label{wenzel2}
\end{equation}
assuming $\theta_\pi$ is small. Here we have included a factor $\sqrt{2}$ so that $a$ measures the root mean square width of the wall. We shall refer to equations (\ref{wenzel1},\ref{wenzel2}) as Wenzel's result although of course Wenzel was unaware of the possibility of a wetting phase transition. Thus it appears to be possible to induce wetting by increasing the roughness of the substrate although this becomes increasingly more difficult for planar contact angles close to $\frac{\pi}{2}$. However one should be suspicious of this prediction given that this approach makes no mention of the order of the wetting transition occuring in the planar system (at temperature $T_\pi$ say). In particular, experience with the well developed finite-size scaling theory of bulk critical phenomena suggests that a more reliable expression for $r_W$ would contain information about the (transverse) correlation length, which recall diverges at a second order wetting transition. Similarly at this simple level we are not able to offer any prediction whether the order of the wetting transition in the non-planar system is different to that occuring for planar geometry. In fact, as we shall see the Wenzel result is inaccurate as regards the influence of roughness on second-order wetting transitions which turns out to be much more interesting than (\ref{wenzel2}) suggests. However, we are able to provide a microscopic derivation of (\ref{wenzel1}) and(\ref{wenzel2}) for first order wetting. 

Before we discuss the microscopic Landau theory we note that it is straight forward to develop a scaling theory for the roughness induced phase shift which improves on (\ref{wenzel2}). Let us suppose that in the planar system the wall-$\alpha$ interface undergoes a continuous (second-order) wetting transition as $t=\frac{T_\pi-T}{T_\pi} \rightarrow 0$. By definition the contact angle $\theta_\pi$ vanishes as $\theta_\pi \sim t^{\frac{2-\alpha_S}{2}}$ where $\alpha_S$ is the specific heat exponent characterising the singular part of the excess free energy $f_{\rm sing} \sim t^{2-\alpha_S}$. Assuming that the wall has the corrugated shape mentioned above it is natural to expect that the length scales $a$ and $q$ enter the singular part of the non-planar free energy through scaled variables $a t^{\beta_S}$ and $qt^{-\nu_\parallel}$ where $\beta_S$ and $\nu_\parallel$ are the adsorption and transverse correlation length critical exponents respectively \cite{schick}. In this way we deduce that the phase boundary in the non-planar geometry (implicitly incorporating the wetting temperature shift) satisfies the scaling relation 
\begin{equation}
\theta_\pi(t,\ldots) \sim q^{\frac{d-1}{2}} S(aq^{\frac{3-d}{2}}) \label{scale}
\end{equation}
where we have used the hyperscaling relations $2-\alpha_S=(d-1)\nu_\parallel$ and $\beta_S=\frac{3-d}{2} \nu_\parallel$ valid below (and at) the upper critical dimension $d^*$ (which recall is equal to three for systems with short range forces \cite{schick}). Here $S(w)$ is an unknown scaling function which must satisfy $S(0)=0$ in order to reproduce the trivial planar phase boundary $\theta_\pi=0$. Interestingly Wenzel's result (\ref{wenzel2}) does meet with the scaling requirement (\ref{scale}) although as we shall see the correct mean-field scaling function $S(w)$ is more complicated and shows non-analytic behaviour. With these preliminaries in mind we now turn to the analysis of a microscopic density-functional model.

Consider a one component system with order parameter $m({\bf r})$ which shows bulk two-phase coexistence at sub-critical temperatures $(T<T_C)$ in zero bulk field $h=0$ between phases with order parameters $m_\alpha(T) \; (>0)$ and $m_\beta(T) \; (<0)$ respectively. If the system is bounded by a fixed wall whose position is specified by a height variable $z_W({\bf r}_\parallel)$ we suppose that the free energy functional $F[m({\bf r})]$ accounting for bulk and wall interactions is
\begin{eqnarray}
F[m({\bf r})] & =& \int d{\bf r} \left\{ \frac{1}{2} (\nabla m)^2 + \phi(m) \right. \nonumber \\
& & \left. + \delta(z-z_W({\bf r}_\parallel)) [1+\frac{1}{2} (\nabla z_W)^2] \phi_1(m) \right\} \label{fe}
\end{eqnarray}
which naturally generalizes the standard free energy model of the planar semi-infinite system \cite{fisher}. As usual the bulk free energy term $\phi(m)$ has a double well form for $T<T_C$ but will not be specified further. The surface interaction term $\phi_1(m)$ \cite{fisher} is taken to have the standard expression $\phi_1(m)=\frac{c m^2}{2} - h_1 m$ where $c$ is the surface enhancement and $h_1$ is the surface field. Note that the surface term contains an extra factor $1+\frac{1}{2}(\nabla z_W)^2$ which accounts for the increase in surface area due to deviations in the position of the wall from the plane \cite{rejmer}. We will always assume that these deviations are small and also of long wavelength compared to some appropriate microscopic scale (see later).

It is natural to anticipate that for small deviations from the plane the minimum of $F[m({\bf r})]$ may be written as a perturbation about the planar value
\begin{eqnarray}
F_\Xi & = & \phi(m_b)V+ \sigma A_\pi \nonumber \\
& & + \frac{1}{2 (2\pi)^{d-1}} \int d{\bf q} \; q^2 \Delta_\pi(q) |\hat{z}_W({\bf q})|^2 + \cdots \label{ansatz}
\end{eqnarray}
where the ellipses denote higher order products of the Fourier amplitudes $\hat{z}_W({\bf q})$ which can be safely ignored for small deviations. The first two terms represent the free energy of the planar system so that $\phi(m_b)$ and $\sigma$ correspond to the bulk free energy density and surface tension respectively. The latter quantity is given by the well known expression \cite{RW,schick,fisher,sullivan}
\begin{equation}
\sigma=\phi_1(m_1) \pm \int_{m_1}^{m_b} dm Q_0(m) \label{sigma}
\end{equation}
where $m_b$ and $m_1$ are bulk and surface magnetizations respectively, and
\begin{equation}
Q_0(m)=\sqrt{2[\phi(m)-\phi(m_b)]}
\end{equation}
The sign in (\ref{sigma}) is chosen so that the contribution from the integral is positive. This expression is amenable to a well known graphical interpretation which is extremely useful in determining the (planar) surface phase diagram \cite{sullivan}.

The quantity $\Delta_\pi(q)$ to be determined represents the free energy correction due to roughness and has the same dimensions as the surface tension $\sigma$. Note that if $\Delta_\pi(q)$ were wave vector independent then the free energy increment is simply proportional to the increase in surface area $A_\Xi-A_\pi$. 

The starting point in our analysis is an exact linear response relation for $\Delta_\pi(q)$ in terms of the (Fourier transformed) planar pair correlation function $G({\bf r}_1,{\bf r}_2)=<m({\bf r}_1)m({\bf r}_2)>-<m({\bf r}_1)><m({\bf r}_2)>$ where both particles are exactly at the wall:
\begin{equation}
G(0,0;{\bf q})=\int d{\bf r}_\parallel \; {\rm e}^{i {\bf q}.{\bf r_\parallel}} G({\bf r}_1,{\bf r}_2) 
\end{equation}
with ${\bf r}_1=(0,0)$ and ${\bf r}_2=(0,{\bf r}_\parallel)$. The result is conveniently written 
\begin{equation}
q^2 \Delta_\pi(q) = q^2 \phi_1(m_1) + {m'}_1^2 \left( \frac{1}{G(0,0;{\bf q})}-\frac{1}{\chi_{11}} \right) \label{delta}
\end{equation}
where $m'_1$ is the derivative of the (planar) wall magnetization with respect to $z$ and $\chi_{11}$ is the surface susceptibility $\chi_{11} \equiv \frac{\partial m_1}{\partial h_1}$. While a derivation of (\ref{delta}) is not appropriate here \cite{us} some remarks should clarify its origin. The first term is the `direct' potential contribution due to the increase in surface area. Note also that the term in parentheses vanishes when $q=0$ (by virtue of a sum rule) reflecting the arbitrariness in the location of the $z=0$ plane. The all important dependence on $G(0,0;{\bf q})$ follows from remarks made earlier by one of us \cite{PARRY} concerning the exact relationship between $G(z,z;{\bf q})$ and the free energy cost of a fluctuation in the location of a surface of (appropriately) fixed magnetization $m^X$ whose average position is $z$. Fortunately a good deal of information about $G(0,0;{\bf q})$ at wetting transitions is now known so that (\ref{delta}) constitutes a rather useful relation. In fact it is possible to continue further with the analysis for {\it arbitrary} $\phi(m)$ (and $\phi_1(m)$) and derive an elegant expression for $\Delta_\pi(q)$ comparable with the surface tension formula (\ref{sigma}). To proceed we substitute into (\ref{delta}) the known, exact expressions \cite{PARRY2} for the moments $G_{2n}(0,0)$ appearing in the expansion $G(0,0;{\bf q})=\sum_{n=0}^{\infty} q^{2n} G_{2n}(0,0)$. While these moments are themselves rather cumbersome the corresponding moments $\Delta_{2n}$ in the expansion $\Delta_\pi(q)=\sum_{n=0}^\infty q^{2n} \Delta_{2n}$ turn out to be much more compact. In this way we we arrive at the following equation for $\Delta_\pi(q)$ valid for arbitrary $\phi(m)$ and $\phi_1(m)$
\begin{equation}
\Delta_\pi(q)=\phi_1(m_1) \pm \int_{m_1}^{m_b} dm Q(m;m_1,q) \label{delta2}
\end{equation}
where the sign is the same as that in (\ref{sigma}). Here the new function $Q(m;m_1,q)$ satisfies the integral equation
\begin{eqnarray}
& \frac{Q(m;m_1,q)}{Q_0(m)} =  & \nonumber \\
&  1-q^2\int_{m_1}^{m} dm' Q_0^{-3}(m') \int_{m'}^{m_b} dm'' Q(m'';m_1,q) & \label{q} 
\end{eqnarray}
where we have assumed that $m_b>m_1$ so that the sign of the double integral is positive. The function $Q(m;m_1,q)$ satisfies the boundary conditions $Q(m_b;m_1,q)=0$ and $Q(m_1;m_1,q)=Q_0(m_1)$ which are useful when interpretating the equation graphically. An important result immediately follows from (\ref{sigma}), (\ref{delta2}) and (\ref{q}) namely
\begin{equation}
\Delta_\pi(0)=\sigma \label{theend}
\end{equation}

\begin{figure}
\epsfxsize=7cm
\vspace{2.5cm}
\epsffile{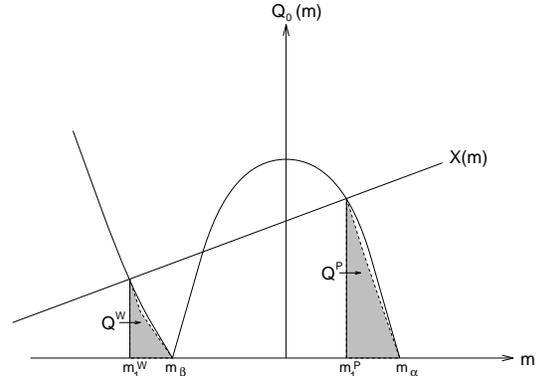}
\caption{Illustration of the generalized graphical construction method for first-order wetting. The dashed lines represent the non-planar $Q$ functions while the shaded areas show the contributions to the free energy from the second term in (11) for wet (W) and partially wet (P) profiles.
}
\end{figure}

This identity reflects the invariance of the free energy of a planar wall-fluid interface with respect to rotations and is related to the asymptotic coherence of surface correlations discussed at length in \cite{parry}. We have also found it profitable \cite{us} to consider the differential version of (\ref{q}) conveniently expressed in terms of the dimensionless scaling factor $y(m;m_1,q) \equiv \frac{Q(m;m_1,q)}{Q_0(m)}$ which satisfies the boundary conditions $y(m_1;m_1,q)=1$ and $y(m_b;m_1,q)=0$
\begin{equation}
Q_0^2(m) y'' + 3Q_0(m) Q_0'(m) y' - q^2 y=0
\end{equation}
where prime denotes differentiation with respect to $m$. For the standard `$\phi^4$' theory (in zero bulk field $h=0$) it transpires that $y$ has a scaling structure $y(m;m_1,q) = Y^{(q \xi_b)}(\frac{m}{m_b};\frac{m_1}{m_b})$ where $\xi_b \equiv \kappa^{-1}$ is the bulk correlation length. The function $Y^{(\alpha)}(t;t_1)$ satisfies
\begin{equation}
(1-t^2) \ddot{Y}^{(\alpha)} -6t \dot{Y}^{(\alpha)}- \frac{4 \alpha^2}{1-t^2} Y^{(\alpha)}=0 \label{latta}
\end{equation}
which is an adaptation of Latta's generalized Mathieu equation \cite{latta}. Here the overdot implies differentiation with respect to $t$. Equations (\ref{delta}-\ref{latta}) constitute the main results of our perturbation theory which we now apply to the problem of wetting at a non-planar wall.

An important result follows directly from (\ref{delta}). Assuming that the wetting transition occuring in the planar geometry is second-order it follows that the correlation function $G(0,0;q)$ and hence the non-planar free energy cannot exhibit singular behaviour away from $T_\pi$. Consequently the wetting transition is shifted only if the transition is roughness induced first-order. Similarly a first-order wetting transition in the planar geometry cannot be roughness induced second-order since this would require an unphysical singularity in $G(0,0;q)$. With these preliminary remarks in mind we consider the case of the influence of roughness on strongly first-order and second-order wetting transitions seperately.
\paragraph{Strongly first-order}
In figure 1 we illustrate a useful graphical interpretation of (\ref{q}) for $\Delta_\pi(q)$. Intersections of the straight line $X(m)=cm-h_1$ with $Q_0(m)$ determine the surface magnetizations $m_1^W, m_1^P$ of the wet and partially wet planar magnetization profiles respectively. Thus the non-wet profile starts at $m_1^P$ and increases to $m_\alpha$ while the wet profile starts at $m_1^W$ and increases to $m_\beta$ (with a $\beta \alpha$ interface at infinity).

The function $y(m;m_1,q)$ can be found accordingly for these profiles. It is very nearly unity (provided $q\xi_b \ll 1$) for the non-wet profile and also for the wet profile between $m_1^W$ and $m_\beta$ while it is zero between $m_\beta$ and $m_\alpha$. The corresponding $Q$ functions are also illustrated in figure 1 (in an obvious notation) together with their respective integral contribution to $\Delta_\pi(q)$ shown as shaded areas. In this way we conclude that the correction term $\Delta_\pi^P$ for the partially wet profile is $\Delta_\pi^P \simeq \sigma_{w\alpha}$ while for the completely wet profile it is $\Delta_\pi^W \simeq \sigma_{w\beta}$. Note that the surface tensions of these profiles are given by $\sigma^P=\sigma_{w\alpha}$ and $\sigma^W=\sigma_{w\beta}+\sigma_{\alpha \beta}$ respectively. Then a simple free energy balancing argument shows that the phase boundary is shifted exactly according to Wenzel's result (\ref{wenzel1})

\paragraph{Second-order wetting}
For this case the analysis is more involved due to the presence of long wavelength fluctuations in the adsorbed fluid film. Consequently we only quote our main results \cite{us} which, for simplicity are restricted to the case where the wall is taken to be at position $z_W = \sqrt{2} a \sin qx$ (and recall $q \xi_b \ll 1$). For fixed $\theta_\pi \sim T_\pi-T$ the planar second-order transition is roughness induced first-order for roughness parameter $r$  satisfying
\begin{equation}
r_W \approx \sec \theta_\pi +(\sec \theta_\pi-1) q^2 \left( \frac{c-\kappa}{c+\kappa} \right) \xi_\parallel^2 \label{ele}
\end{equation}
where $\xi_\parallel$ is the planar value of the transverse correlation length at this temperature. Clearly this is a fluctuation modified version of Wenzel's result (\ref{wenzel1}).

\begin{figure}
\epsfxsize=7cm
\epsffile{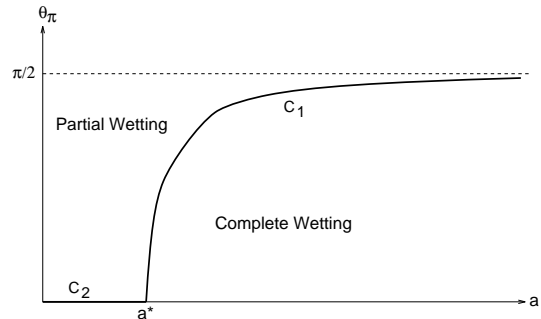}
\caption{Schematic wetting phase diagram for fluid adsorption in a system with a non-planar wall. The lines $C_1$ and $C_2$ are loci of first and second-order wetting phase transitions respectively which meet at a tricritical point corresponding to $a=a^* \leq \xi_b$. The vertical axis is a linear measure of the temperature scale $T_\pi -T$ for small $\theta_\pi$.}
\end{figure}

Note the factor $\frac{c-\kappa}{c+\kappa} >0$ is a measure of the deviation from the planar tricritical point. For small values of $\theta_\pi$ i.\ e.\ $T$ close to $T_\pi$ the result reduces to 
\begin{equation}
r_W \simeq \sec \theta_\pi + R \; (q \xi_b)^2 \left( \frac{c-\kappa}{c+\kappa} \right)
\end{equation}
where
\begin{equation}
R=\stackrel{\rm lim}{\scriptscriptstyle{T \rightarrow T_\pi}} \left\{ \frac{f_{\rm sing} \xi_\parallel^2}{\sigma_{\alpha \beta} \xi_b^2} \right\}
\end{equation}
is the ratio of two hyperscaling amplitudes and takes the universal value $R=\frac{1}{2}$ in mean field theory. Thus the shifted phase boundary (for $\theta_\pi \gtrsim 0$) is given by the non-analytic function
\begin{equation}
\theta_\pi = \left\{ \begin{array}{l} 0 \mbox{\ \ \ \ for  $a<a^*$} \\ q(a^2-a^{*2})^{\frac{1}{2}} \mbox{\ \ \ \ for  $a>a^*$} \end{array} \right.
\end{equation}
where 
\begin{equation}
a^*= \sqrt{ \frac{c-\kappa}{c+\kappa} } \xi_b \label{rt}
\end{equation}
is the tricritical value of $a$. The analytical expression is confirmed by numerical minimization of the mean field free energy functional \cite{us}. Thus for fixed $a<a^*$ the wetting phase transition remains second-order and occurs at the (unshifted) temperature $T_\pi$. While for $a>a^*$ the transition is roughness induced first-order and occurs at a {\it lower} temperature. The phase diagram plotted in terms of $\theta_\pi$ and $a$ is sketched in figure 2. Importantly, the expression for the shifted phase boundary is precisely of the scaling form (\ref{scale}) provided we identify $d=3$ corresponding to the upper critical dimension for wetting (for models with short ranged forces) as we might anticipate for the present mean field calculation. Interestingly as $c$ is reduced to $\kappa$ (i.\ e.\ as we approach the planar tricritical point) we recover the simple prediction of Wenzel (\ref{wenzel2}). It is also noteworthy that the tricritical value of the width parameter $a^*$ given by (\ref{rt}) is independent of $q$ (for $q \xi_b \ll 1$). The fact that a second order wetting transition is roughness induced first-order for even minor deviations from the plane is the central conclusion of our study. Moreover experience with (improved) effective Hamiltonian models of wetting in three dimensional systems is strongly suggestive that this result is true beyond mean field theory since there does not appear to be a mechanism (involving a position dependent stiffness or stiffness matrix \cite{parry2}) by which a first-order wetting transition is fluctuation induced second-order. Thus we are confident that the topology of the surface phase diagram (see figure 2) would be borne out in Ising model simulation studies.

In summary, we have developed a mean field linear response theory for fluid adsorption at a non-planar wall and derived an exact analytical expression for the perturbation of the planar free energy due to surface roughness. Using this method we have been able to vindicate a phenomenological result (which may be attributed to Wenzel) regarding the influence of roughness on first-order wetting transitions. For second-order wetting transitions however, Wenzel's result is inappropriate (due to fluctuation effects) and calculation shows that the transition is driven first-order for deviations from the plane.

We are very grateful for Professor M.\ Napi\'{o}rkowski for discussions which reawakened our interest in this problem and acknowledge financial support from the E.\ P.\ S.\ R.\ C.\ (United Kingdom).

\end{document}